\documentclass[
reprint,
 amsmath,amssymb,
 aps,
 prl,
]{revtex4-2}

\usepackage{graphicx}
\usepackage{dcolumn}
\usepackage{bm}
\usepackage{hyperref}
\usepackage[utf8]{inputenc}

\newcommand{\ket}[1]{\left| #1 \right\rangle}
\newcommand{\braket}[2]{\langle #1 | #2 \rangle}

\renewcommand{\Re}{\mathrm{Re}}
\renewcommand{\Im}{\mathrm{Im}}

\newcommand{\sn}{\mathrm{sn}}
\newcommand{\cn}{\mathrm{cn}}
\newcommand{\dn}{\mathrm{dn}}

\begin{document}


\title{Interference and Oscillation in Nambu Quantum Mechanics}

\author{Djordje Minic}
\email{dminic@vt.edu}

\author{Tatsu Takeuchi}
\email{takeuchi@vt.edu}

\author{Chia Hsiung Tze}
\email{ctze@me.com}

\affiliation{Center for Neutrino Physics, Department of Physics, Virginia Tech, Blacksburg VA 24061, USA}

\begin{abstract}
Nambu Quantum Mechanics, proposed in Phys. Lett. B536, 305 (2002),
is a deformation of canonical Quantum Mechanics in which
only the time-evolution of the ``phases'' of energy eigenstates is modified.
We discuss the effect this theory will have on oscillation phenomena,
and place a bound on the deformation parameters utilizing the data on the atmospheric neutrino mixing angle $\theta_{23}$.
\end{abstract}


\maketitle

\noindent
{\bf Introduction:} Quantum Mechanics (QM) is one of the most important and successful frameworks of modern physics.
The language of QM is essential for particle, nuclear, atomic, condensed matter, and statistical physics
as well as chemistry, 
and it has lead to the current ``second quantum revolution'' in quantum information science and technology 
\cite{PRXQuantum.1.020101}.
Nevertheless, the full understanding of the foundations and origins of QM is still an 
active area of intense discussion and research \cite{Bell:1987hh,Aharonov:2005,Hooft:2014kka}. 
It has been argued that canonical QM should be replaced by a more fundamental or generalized framework, 
either in the context of quantum gravity and cosmology \cite{Penrose:2014nha,Gell-Mann:2013hza}, 
or in the realm of quantum measurement \cite{Weinberg:2016axv}, 
or in the domain of macroscopic quantum systems \cite{Leggett:2007zz}. 

A deeper understanding of canonical QM could be obtained by comparing its
predictions to those of its possible generalizations, and confronting both with experiment.
It would allow us to probe the robustness of the original tenet
or axiom that was relaxed to generalize the theory,
thereby identify the theoretical bedrock on which QM rests.
(See e.g. \cite{Raghavan:2012sy} and references therein).

Various proposals for alternative QM theories can be found in the literature.
The field/division algebra over which the state space is constructed
has been modified from $\mathbb{C}$ to 
$\mathbb{R}$ \cite{Stueckelberg:1960}, 
$\mathbb{H}$ \cite{Adler:1988hb}, 
$\mathbb{O}$ \cite{Gunaydin:1978jq,Okubo:1990nv,Gursey:1996mj}, and the finite fields
$\mathbb{F}_q$ \cite{Chang:2012eh,Chang:2012gg,Chang:2012we,Takeuchi:2012mra,Chang:2013joa,Chang:2013rya,Chang:2019kcp}.
Non-linear corrections to the Schr\"odinger equation have been considered by Weinberg \cite{Weinberg:1989cm,Weinberg:1989us}.
Other ideas address issues in quantum measurement and quantum gravity
\cite{Penrose:2014nha,Gell-Mann:2013hza,Weinberg:2016axv}.

In this letter, we look at Nambu QM introduced in \cite{Minic:2002pd} by Minic and Tze,
and one of its consequences.
The work was inspired by a profound and far-reaching paper by Nambu \cite{Nambu:1973qe}.
The starting point of the approach is the geometric formulation of QM \cite{Kibble:1978tm}
in which the time evolution of pure quantum states is described as 
a ``classical'' area preserving Hamiltonian flow within the state ``phase'' space.
For a single energy eigenstate, this is just the evolution of its phase, the
real and imaginary parts of which constitute the ``phase'' space, with the Hamiltonian
being that of a harmonic oscillator.
Nambu's idea in \cite{Nambu:1973qe} was to extend the classical equation of motion $\dot{F} = -\{H,F\}$, where
\begin{equation}
\{A,B\} \;=\; \varepsilon_{ij}\dfrac{\partial A}{\partial q_i}\dfrac{\partial B}{\partial q_j}
\label{PoissonBracket}
\end{equation}
is the Poisson bracket, to $\dot{F} = -\{H_1,H_2,F\}$, where
\begin{equation}
\{A,B,C\} \;=\; \varepsilon_{ijk}\dfrac{\partial A}{\partial q_i}\dfrac{\partial B}{\partial q_j}\dfrac{\partial C}{\partial q_k}\;,
\label{NambuBracket}
\end{equation}
\textit{i.e.} the Nambu bracket.
In Poisson dynamics time evolution is generated by the one conserved quantity $H$,
while Nambu dynamics requires two: $H_1$ and $H_2$, and the generated flow is volume preserving.
An application of the Nambu equation is the asymmetric top in which the evolution of its
angular momentum $\vec{L}$ can be generated by the Nambu bracket with the energy $E$ and 
total angular momentum $L^2/2$ serving the roles of $H_1$ and $H_2$.
The proposal of \cite{Minic:2002pd} was to enlarge the ``phase'' space of each
energy eigenstate from two dimensions to three, and 
assume that the ``classical motion'' of the ``phase'' was governed by 
Nambu asymmetric top dynamics instead of that of a Poisson harmonic oscillator.
Note that this deformation of canonical QM is particularly
attractive since it is minimalistic: it only deforms the time-evolution of
the ``phase'' of energy eigenstates and everything else is kept fixed.
Furthermore, this deformation can be continuously turned on and off.

In the following, we first review canonical QM in the two-component real vector notation 
and the treatment of oscillations in that language. 
Then, we formulate Nambu QM as the three-component real vector extension 
to canonical QM
and derive an explicit formula for oscillations in this context. 
This formula can be probed experimentally,
most promisingly in neutrino oscillations.

\smallskip
\noindent
{\bf Canonical QM:}
Let $\ket{n}$ denote the $n$-th eigenstate of the Hamiltonian $\hat{H}$ with
eigenvalue $E_n=\hbar\omega_n$ :
\begin{equation}
\hat{H}\ket{n} \;=\; E_n\ket{n}\;.
\end{equation}
A generic state $\ket{\psi(t)}$ can be expanded in terms of $\ket{n}$ as 
\begin{equation}
\ket{\psi(t)} 
\;=\; \sum_{n} \ket{n}\underbrace{\braket{n}{\psi(t)}}_{\displaystyle \psi_n(t)} 
\;=\; \sum_{n} \psi_n(t)\ket{n}\;,
\end{equation}
where the coefficients $\psi_n(t)$ evolve in time as
\begin{equation}
\psi_n(t) \;=\; N_n\,e^{-i\omega_n(t-t_n)}\;.
\end{equation}
Here, we take $N_n$ to be real and positive, and $t_n$ is the boundary time at which $\psi_n(t)$ is phaseless.
The complex valued $\psi_n(t)$ can also be expressed as a two-component real vector as
\begin{equation}
\vec{\psi}_n(t) \;\equiv\;
\begin{bmatrix} \Re\,\psi_n(t) \\ \Im\,\psi_n(t) \end{bmatrix}
\;=\;
N_n
\begin{bmatrix}
\phantom{-}\cos\omega_n(t-t_{n}) \\
-\sin\omega_n(t-t_{n})
\end{bmatrix}
\;.
\label{TimeEvolution2}
\end{equation}
The inner product between two states $\ket{\psi}$ and $\ket{\phi}$ in this
two-component real vector notation is
\begin{equation}
\!\!\!\!\braket{\psi}{\phi} 
= \sum_{n}\psi_n^*\phi_n 
=
\underbrace{\sum_{n}(\vec{\psi}_n\cdot\vec{\phi}_n)}_{\displaystyle g(\psi,\phi)} 
+ i\underbrace{\sum_{n}(\vec{\psi}_n\times\vec{\phi}_n)}_{\displaystyle \varepsilon(\psi,\phi)}
\;.
\end{equation}
Note that $g(\psi,\phi)$ and $\varepsilon(\psi,\phi)$ 
depend only on the magnitudes of, and the relative angles between the
$(\vec{\psi}_n,\vec{\phi}_n)$ pairs. 
They are invariant under 2D rotations.
The absolute value of the inner product squared is then
\begin{equation}
|\braket{\psi}{\phi}|^2 
\;=\; g(\psi,\phi)^2 + \varepsilon(\psi,\phi)^2\;.
\end{equation}

Now, consider two energy eigenstates $\ket{1}$ and $\ket{2}$ and two flavor eigenstates $\ket{\alpha}$ and $\ket{\beta}$
which are related by
\begin{equation}
\begin{bmatrix} \ket{\alpha} \\ \ket{\beta} \end{bmatrix}
\;=\; 
\begin{bmatrix} c_\theta & s_\theta \\ -s_\theta & c_\theta \end{bmatrix}
\begin{bmatrix} \ket{1} \\ \ket{2} \end{bmatrix}
\;,
\label{mixing}
\end{equation}
where $s_\theta=\sin\theta$, $c_\theta=\cos\theta$.
In vector notation, we have
\begin{equation}
\begin{array}{rlrl}
  \vec{\alpha}_1 & =\; c_\theta\,\vec{n}_0\;,\qquad\qquad
& \vec{\alpha}_2 & =\; s_\theta\,\vec{n}_0\;,
\\
  \vec{\beta}_1 & =\; -s_\theta\,\vec{n}_0\;,\qquad\qquad
& \vec{\beta}_2 & =\;  c_\theta\,\vec{n}_0\;,
\end{array}
\label{alpha_beta_n0}
\end{equation}
where $\vec{n}_0$ represents a phaseless state:
\begin{equation}
\vec{n}_0 \;=\; \begin{bmatrix} 1 \\ 0 \end{bmatrix}\;.
\end{equation}
Let $\ket{\psi(0)}=\ket{\alpha}$, that is
\begin{equation}
\vec{\psi}_1(0) = \vec{\alpha}_1 = c_\theta\,\vec{n}_0\;,\qquad
\vec{\psi}_2(0) = \vec{\alpha}_2 = s_\theta\,\vec{n}_0\;.
\end{equation}
At a later time, these will have evolved into
\begin{equation}
\vec{\psi}_1(t)
= c_\theta\,\vec{n}_1(t)\;,\qquad
\vec{\psi}_2(t)
= s_\theta\,\vec{n}_2(t)\;,
\end{equation}
where
\begin{equation}
\vec{n}_1(t) \;=\; \begin{bmatrix} c_1 \\ -s_1 \end{bmatrix}\;,\qquad
\vec{n}_2(t) \;=\; \begin{bmatrix} c_2 \\ -s_2 \end{bmatrix}\;,
\end{equation}
with $s_i = \sin\omega_i t$ and $c_i = \cos\omega_i t$.
To find the survival probability $P(\alpha\to\alpha)$ of flavor $\alpha$,
and the transition probability $P(\alpha\to\beta)$ to flavor $\beta$, we need 
$\braket{\alpha}{\psi(t)}$ and $\braket{\beta}{\psi(t)}$.
The symmetric and antisymmetric parts of these inner products are
\begin{eqnarray}
g(\alpha,\psi(t)) & = &
 \vec{\alpha}_1\cdot\vec{\psi}_1(t)
+\vec{\alpha}_2\cdot\vec{\psi}_2(t)
= c_\theta^2 c_1 + s_\theta^2 c_2 \;,
\\
\varepsilon(\alpha,\psi(t)) & = & 
 \vec{\alpha}_1\times\vec{\psi}_1(t)
+\vec{\alpha}_2\times\vec{\psi}_2(t)
= -c_\theta^2 s_1 - s_\theta^2 s_2 \;,
\vphantom{\Big|}\cr
g(\beta,\psi(t)) & = &
 \vec{\beta}_1\cdot\vec{\psi}_1(t)
+\vec{\beta}_2\cdot\vec{\psi}_2(t)
= -s_\theta c_\theta c_1 + s_\theta c_\theta c_2 \;,
\vphantom{\Big|}\cr
\varepsilon(\beta,\psi(t)) & = & 
 \vec{\beta}_1\times\vec{\psi}_1(t)
+\vec{\beta}_2\times\vec{\psi}_2(t)
= s_\theta c_\theta s_1 - s_\theta c_\theta s_2 \;,
\vphantom{\Big|}\nonumber
\end{eqnarray}
and the survival and transition probabilities will be
\begin{eqnarray}
P(\alpha\to\alpha)
& = & |\braket{\alpha}{\psi(t)}|^2 
\,=\, g(\alpha,\psi(t))^2 + \varepsilon(\alpha,\psi(t))^2 
\vphantom{\Big|}\cr
& = &
1 - P(\alpha\to\beta)\;,
\vphantom{\Big|}\cr
P(\alpha\to\beta)
& = & |\braket{\beta}{\psi(t)}|^2 
\,=\, g(\beta,\psi(t))^2 + \varepsilon(\beta,\psi(t))^2 
\vphantom{\Big|}\cr
& = & \sin^2 2\theta\, \sin^2\!\left[\dfrac{(\omega_1-\omega_2)t}{2}\right]\;.
\vphantom{\bigg|}
\label{PAAandPAB}
\end{eqnarray}
Making the relativistic replacement
\begin{equation}
\omega_i t \;\to\; (\omega_i t - k_i L) \;\xrightarrow{\mathrm{natural\ units}}\;
(E_i t - p_i L)\;,
\label{Replacement1}
\end{equation}
and assuming that the energies of the two
states are common, $E_1=E_2=E \gg m_{i}$, we have
\begin{equation}
(E t - p_i L)
\;\approx\;
E(t-L) + \dfrac{m_i^2}{2E}L\;,
\label{Replacement2}
\end{equation}
leading to the identification
\begin{equation}
(\omega_1-\omega_2)t
\;\;\to\;\; \dfrac{\delta m_{12}^2}{2E}\,L \;\equiv\; \Delta_{12}\;.
\label{Replacement3}
\end{equation}
This gives us the familiar neutrino oscillation formula
\begin{equation}
P(\alpha\to\beta) \;=\;
\sin^2 2\theta\;\sin^2\dfrac{\Delta_{12}}{2}\;.
\label{CanonicalOscillation}
\end{equation}

\noindent
{\bf Nambu QM:}
The deformation of canonical QM which is detailed in Ref.~\cite{Minic:2002pd},
\textit{i.e.} Nambu QM, 
can be summarized as follows.
Extend the two-component real vector $\vec{\psi}_n$ introduced above to 
a three-component real vector $\vec{\Psi}_n$:
\begin{equation}
\vec{\psi}_n \quad\to\quad
\vec{\Psi}_n \;.
\end{equation}
In the two-component case, the components evolved as
Eq.~\eqref{TimeEvolution2}.  
For the three-component extension, it is assumed that
\begin{equation}
\vec{\Psi}_n(t) \;=\; N_n
\begin{bmatrix} 
\phantom{-\kappa}c_\xi\,\cn(\Omega_n(t-t_n),k) \\ 
         -\kappa c_\xi\,\sn(\Omega_n(t-t_n),k) \\ 
            \;\;-s_\xi\,\dn(\Omega_n(t-t_n),k) 
\label{TimeEvolution3}
\end{bmatrix}
\;,
\end{equation} 
where $\cn(u,k)$, $\sn(u,k)$, and $\dn(u,k)$ are Jacobi's elliptical functions \footnote{%
Mathematica encodes $\sn(u,k)$, $\cn(u,k)$, and $\dn(u,k)$ respectively as
\texttt{JacobiSN[$u$,$m$]}, \texttt{JacobiCN[$u$,$m$]}, and \texttt{JacobiDN[$u$,$m$]} with $m=k^2$.
},
and $s_\xi = \sin\xi$, $c_\xi = \cos\xi$, and $\kappa = \sqrt{1+k^2\tan^2\xi}$.
The period of $\cn(u,k)$ and $\sn(u,k)$ in $u$ is $4K$, where $K=K(k)$ is the
complete elliptical integral of the first kind \footnote{%
Mathematica encodes $K(k)$ as \texttt{EllipticK[$m$]} with $m=k^2$.}, 
and $\Omega_n = (2K/\pi)\omega_n$.
The two parameters $k$ and $\xi$ are the deformation parameters,
and when they are both set to zero, the time evolution of the first two components
of $\vec{\Psi}_n$ reduce to that of the two components of $\vec{\psi}_n$,
while the third component of $\vec{\Psi}_n$ vanishes.
In principle, we can make the deformation parameters $k$ and $\xi$ depend on $n$,
but for the sake of simplicity, we keep them common to all $n$.

Note that the time evolution assumed in Eq.~\eqref{TimeEvolution3}
is that of the angular momentum vector $\vec{L}$ of a free asymmetric top
in its co-rotating frame \cite{Opatrny:2018}.
Though the equations that govern this motion are \textit{non-linear} (or more precisely \textit{multi-linear}), 
the presence of the two conserved quantities of energy $E$ and angular momentum $L^2$
renders the motion solvable, norm-preserving, and periodic.
$\vec{L}$ evolves along the intersection of the sphere $L^2=\mathrm{constant}$ and the
ellipsoid $E=\mathrm{constant}$.
Due to the norm preserving nature of Eq.~\eqref{TimeEvolution3},
this time evolution is \textit{unitary}. 
However, the time evolution operator cannot be expressed as a matrix
as in canonical QM (except when $k=0$) due to the evolution being non-linear.
In essence, the ``phase'' of the state evolves periodically on $S^2$ instead of on $S^1$.

The symmetric and antisymmetric parts of the inner product between two states are
extended to
\begin{eqnarray}
g(\Psi,\Phi) & = & \sum_{n}(\vec{\Psi}_n\cdot\vec{\Phi}_n)\;,\vphantom{\Big|}\cr
\vec{\varepsilon}\,(\Psi,\Phi) & = & \sum_{n}(\vec{\Psi}_n\times\vec{\Phi}_n)\;,\vphantom{\Big|}
\end{eqnarray}
where the dot and cross products are now defined in three dimensions. 
Consequently, $\vec{\varepsilon}\,$ has three components, which in the $k=\xi=0$ limit
reduces to
\begin{equation}
\sum_{n}(\vec{\Psi}_n\times\vec{\Phi}_n)
\;\xrightarrow{k=\xi=0}\;
\begin{bmatrix} 
0\vphantom{\big|} \\ 
0\vphantom{\big|} \\ 
\sum_n(\vec{\psi}_n\times\vec{\phi}_n)\vphantom{\Big|}
\end{bmatrix}
\;.
\end{equation}
The square of the absolute value of $\braket{\Psi}{\Phi}$ is extended to
\begin{equation}
|\braket{\Psi}{\Phi}|^2 \;=\;
g(\Psi,\Phi)^2 + \vec{\varepsilon}\,(\Psi,\Phi)\cdot\vec{\varepsilon}\,(\Psi,\Phi)\;,
\end{equation}
which is invariant under 3D rotations of the ``phase'' space.
This expression allows us to make predictions based on Nambu QM.
Since the deformation is in the time-evolution of the ``phase'' of each energy eigenstate,
we can expect deviations from canonical QM to occur in phenomena
that involve the evolution of interference terms.

Consequently, let us look at oscillation in Nambu QM.
We consider flavor eigenstates $\ket{\alpha}$ and $\ket{\beta}$ to be
superpositions of energy eigenstates $\ket{1}$ and $\ket{2}$ as in Eq.~\eqref{mixing}.
The three-component vector notation of $\ket{\alpha}$ and $\ket{\beta}$ 
are formally the same as Eq.~\eqref{alpha_beta_n0}, except with $\vec{n}_0$
replaced by the three component object
\begin{equation}
\vec{n}_0\;=\;\begin{bmatrix} c_\xi \\ 0 \\ -s_\xi \end{bmatrix}\;.
\label{phaseless}
\end{equation}
This corresponds to a ``zero phase'' state.
To clarify that we are working in the three-component formalism,
we will replace the label $\alpha$ with $A$, and $\beta$ with $B$
in the following.

Let $\ket{\Psi(0)} = \ket{A}$, that is:
\begin{equation}
\vec{\Psi}_1(0) \;=\; c_\theta\,\vec{n}_0\;,\qquad
\vec{\Psi}_2(0) \;=\; s_\theta\,\vec{n}_0\;.
\end{equation}
At a later time $t$, these will evolve to
\begin{equation}
\vec{\Psi}_1(t) \;=\; c_\theta\,\vec{n}_1(t)\;,\qquad
\vec{\Psi}_2(t) \;=\; s_\theta\,\vec{n}_2(t)\;,
\end{equation}
where
\begin{equation}
\vec{n}_i(t) 
\;=\;
\begin{bmatrix}
\phantom{-\kappa}c_{\xi}\, \cn_i \\
-\kappa c_{\xi}\, \sn_i \\
\;\;-s_{\xi}\, \dn_i
\end{bmatrix}\;,
\end{equation}
with $\sn_i = \sn(\Omega_i t,k)$, $\cn_i = \cn(\Omega_i t,k)$, $\dn_i = \dn(\Omega_i t,k)$.
The symmetric parts of $\braket{A}{\Psi(t)}$ and $\braket{B}{\Psi(t)}$ are
\begin{eqnarray}
\lefteqn{g(A,\Psi(t))
\;=\;\vec{A}_1\cdot\vec{\Psi}_1(t) + \vec{A}_2\cdot\vec{\Psi}_2(t)
}
\\
& = & c_\theta^2
\Bigl(\,
c_\xi^2\, \cn_1 + s_\xi^2\, \dn_1
\,\Bigr)
+ s_\theta^2
\Bigl(\,
c_\xi^2\, \cn_2 + s_\xi^2\, \dn_2
\,\Bigr)
\;,\vphantom{\bigg|}\cr
\lefteqn{g(B,\Psi(t))
\;=\; \vec{B}_1\cdot\vec{\Psi}_1(t) + \vec{B}_2\cdot\vec{\Psi}_2(t)
}
\vphantom{\Big|}\cr
& = & -s_\theta c_\theta
\Bigl(\,
c_\xi^2\, \cn_1 + s_\xi^2\, \dn_1
\,\Bigr)
+ s_\theta c_\theta
\Bigl(\,
c_\xi^2\, \cn_2 + s_\xi^2\, \dn_2
\,\Bigr)
\;,\vphantom{\Big|}\nonumber
\end{eqnarray}
while the antisymmetric parts are
\begin{eqnarray}
\lefteqn{\vec{\varepsilon}\,(A,\Psi(t))
\;=\; \vec{A}_1\times\vec{\Psi}_1(t) + \vec{A}_2\times\vec{\Psi}_2(t)
}
\\
& = & c_\theta^2
\begin{bmatrix}
-\kappa s_\xi c_\xi\,\sn_1 \\
s_\xi c_\xi \bigl( \dn_1 - \cn_1 \bigr) \\
-\kappa c_\xi^2\, \sn_1
\end{bmatrix}
+ s_\theta^2
\begin{bmatrix}
-\kappa s_\xi c_\xi\,\sn_2 \\
s_\xi c_\xi \bigl( \dn_2 - \cn_2 \bigr) \\
-\kappa c_\xi^2\, \sn_2
\end{bmatrix}
\;,\cr
& & \vphantom{x}\cr
\lefteqn{\vec{\varepsilon}\,(B,\Psi(t))
\;=\; \vec{B}_1\times\vec{\Psi}_1(t) + \vec{B}_2\times\vec{\Psi}_2(t)
\vphantom{\Big|}}\cr
& = & -s_\theta c_\theta
\begin{bmatrix}
-\kappa s_\xi c_\xi\,\sn_1 \\
s_\xi c_\xi \bigl( \dn_1 - \cn_1 \bigr) \\
-\kappa c_\xi^2\, \sn_1
\end{bmatrix}
+ s_\theta c_\theta
\begin{bmatrix}
-\kappa s_\xi c_\xi\,\sn_2 \\
s_\xi c_\xi \bigl( \dn_2 - \cn_2 \bigr) \\
-\kappa c_\xi^2\, \sn_2
\end{bmatrix}
\;.\nonumber
\end{eqnarray}
From these expressions, we find
the survival and transition probabilities to be
\begin{widetext}
\begin{eqnarray}
P(A\to A) & = & g(A,\Psi(t))^2 + \vec{\varepsilon}\,(A,\Psi(t))\cdot \vec{\varepsilon}\,(A,\Psi(t)) 
\;=\; 1 - P(A\to B)
\;,\vphantom{\Big|}\cr
P(A\to B) & = & g(B,\Psi(t))^2 + \vec{\varepsilon}\,(B,\Psi(t))\cdot \vec{\varepsilon}\,(B,\Psi(t)) 
\vphantom{\bigg|}\cr
& = & 
\sin^2 2\theta
\left[
\dfrac{1
-
\bigl\{c_\xi^2
\bigl( \sn_1\,\sn_2 + \cn_1\,\cn_2 \bigr)
+s_\xi^2
\bigl( k^2\,\sn_1\,\sn_2 + \dn_1\,\dn_2 \bigr)
\bigr\}
}{2}
\right]
\;.
\label{PAAandPAB-NQM}
\end{eqnarray}
\end{widetext}
\noindent
For the ease of comparison with Eq.~\eqref{CanonicalOscillation},
we expand the Jacobi functions in powers of $k^2$ \cite{Gradshteyn-Ryzhik,Abramowitz-Stegun,Whittaker-Watson} :
\begin{eqnarray}
\sn(\Omega t,k) & = & \left(1+\dfrac{k^2}{16}\right)\sin(\omega t) + \dfrac{k^2}{16}\sin(3\omega t) + \cdots\;, \cr
\cn(\Omega t,k) & = & \left(1-\dfrac{k^2}{16}\right)\cos(\omega t) + \dfrac{k^2}{16}\cos(3\omega t) + \cdots\;, \cr
\dn(\Omega t,k) & = & \left(1-\dfrac{k^2}{4}\right) + \dfrac{k^2}{4}\cos(2\omega t) + \cdots\;,
\label{JacobiExpansions}
\end{eqnarray}
from which we find to order $k^2$ 
\begin{eqnarray}
\sn_1\,\sn_2+\cn_1\,\cn_2 
& = & \cos\Delta_{12} - \dfrac{k^2}{4}\cos\Sigma_{12}\sin^2\Delta_{12}  \;,
\cr
k^2\,\sn_1\,\sn_2+\dn_1\,\dn_2 \vphantom{\Big|}
& = & 1 - k^2\big(1 + \cos\Sigma_{12}\big)\sin^2\dfrac{\Delta_{12}}{2} \;,\cr
& &
\end{eqnarray}
where $\Delta_{12}=(\omega_1-\omega_2)t$ and $\Sigma_{12}=(\omega_1+\omega_2)t$.
Averaging over time makes the $\cos\Sigma_{12}$ terms vanish.
Therefore, 
\begin{equation}
P(A\to B) \;=\; 
\left(
 c_\xi^2+s_\xi^2\dfrac{k^2}{2}\right)\sin^2 2\theta\,\sin^2\dfrac{\Delta_{12}}{2}
\;.
\label{MainResult}
\end{equation}
Note that $0\le k^2<1$.
Thus, the effect of the Nambu deformation is an overall suppression factor
compared to the undeformed canonical case, Eq.~\eqref{CanonicalOscillation}.
This is the main result of this letter.

\smallskip
\noindent
{\bf Discussion:}
In this letter, we have considered Nambu QM, a 
deformation of canonical QM in which
the ``phase'' space of energy eigenstates is enlarged from 2D to 3D,
and the ``phase'' dynamics is deformed from that of a harmonic oscillator
to that of an asymmetric top.
This deformation maintains the Born rule, \textit{i.e.} the conservation of norm,
which is embedded in the ``classical'' dynamics of the ``phase''. 
The invariance of physical predictions on 2D rotations of the
``phase'' space is modified to that under 3D rotations,
a feature responsible for the projectivity of the state space.
(Note that we cannot associate a ``phase'' shift with a constant shift of $\Omega t$
in Eq.~\eqref{TimeEvolution3}.)
This invariance can, in principle, be gauged in the field theoretic version
of Nambu QM, but since the symmetry is $SO(3)$, it could lead to
non-abelian features though only the ``phase'' of a single field will be gauged.
The $S^2$ geometry of the ``phase'' space, as opposed to the canonical $S^1$,
also suggests that the path integral of Nambu QM is not the usual 
integral of $e^{iS}$.

We have investigated the effect of the Nambu QM deformation on oscillation phenomena,
relevant \textit{e.g.} for neutrinos and neutral mesons \cite{Kayser:2011jn},
and have derived an explicit formula, Eq.~\eqref{MainResult}, involving the deformation parameters $k$ and $\xi$. 
Given that the $\sin^2 2\theta$ term cannot increase beyond one, the suppression factor
cannot always be absorbed into $\theta$.  For instance, the current bound on the
atmospheric mixing angle gives $\sin^2 2\theta_{23} > 0.973,\;0.963,\;0.952$ 
respectively at 1, 2, and 3$\sigma$ for normal ordering \cite{deSalas:2020pgw}.
This indicates
\begin{equation}
s_\xi^2\left(1-\dfrac{k^2}{2}\right) \;<\; 0.027\;(1\sigma)\;,\; 0.037\;(2\sigma)\;,\; 0.048\;(3\sigma)\;,
\end{equation}
though the value of $\theta_{23}$ itself is not yet precisely known.
Future improvements in the determination of $\theta_{23}$ at
IceCube \cite{Terliuk:2019hkb}, JUNO \cite{Guo:2017mvv}, and DUNE \cite{Higuera:2018zjz}
could improve upon this bound.

Note that oscillation is but one possible phenomenon that could be affected by Nambu QM.
There may be many others involving interference and the resulting correlations
given that the ``phase'' vectors are assumed to move in a very particular way on $S^2$.
They could shed new light on various issues in quantum foundations and in entanglement,
and call for a thorough investigation.

Apart from these phenomenological considerations,
we would like to highlight the fact that the original paper of Nambu \cite{Nambu:1973qe}
has inspired very many works on the mathematical and foundational nature of the Nambu bracket, Eq.~\eqref{NambuBracket}, 
and its related structures \cite{Takhtajan:1993vr,Dito:1996xr,Dito:1996hn}, 
and on the quantization of those structures and their relevance in string theory
(see \cite{Bergshoeff:1988hw,Awata:1997yr,Awata:1999dz,Fujikawa:1997jt,Smolin:1997ai,Curtright:2002fd,Curtright:2002sr,Curtright:2003fn,Bagger:2007jr,Gustavsson:2007vu,Chu:2008qv,Ho:2016hob} and references therein).
More recently, such a structure 
was discovered \cite{Freidel:2019jor,Minic:2020oho,Berglund:2019ctg,Berglund:2020qcu} in the context of a
new formulation of non-perturbative string theory and quantum gravity based
on quantum spacetime \cite{Freidel:2013zga,Freidel:2014qna,Freidel:2015pka,Freidel:2016pls,Freidel:2017xsi,Freidel:2017wst,Freidel:2017nhg,Freidel:2018apz}.

We also note that analogies with the asymmetric top are ubiquitous in various classical and quantum physical systems \cite{Opatrny:2018}. This has particularly been the case in phenomenological particle physics. What we have uncovered here in Nambu QM relates closely to, and formally extends in a new direction the top like Hamiltonians used in dynamical models of neutrino oscillations \cite{Raffelt:2011yb,Pehlivan:2011hp}. 
 They belong to the family of integrable quantum spin Gaudin models of wider applications in condensed matter physics. The time oscillatory features we have deduced in this letter along with the above mentioned connection further suggests the construction and phenomenological testing of a family of dynamical, integrable $SO(3)$ Nambu top models of, say, neutrinos oscillations with, not a trigonometric but a novel Jacobian elliptic time evolutions with two periods -- presumably with one period being much, much smaller than the other. They would add a new prediction for neutrino oscillations in our quest to see theoretically and experimentally beyond the Standard Model.

\smallskip
\noindent
{\bf Acknowlegments:}
We thank P.~Huber and R.~Pestes for helpful discussions.
DM and TT are supported in part by the DOE (DE-SC0020262).
DM is also supported by the Julian Schwinger Foundation, 
and TT by the NSF (PHY-1413031).\\
We dedicate this letter to the memory
of Prof. Yoichiro Nambu, a great physicist and a wonderful human being, on the occasion of
his upcoming centennial in 2021.

\newpage
\onecolumngrid

\bibliographystyle{apsrev4-2}
\bibliography{Nambu}

\begin{thebibliography}{66}%
\makeatletter
\providecommand \@ifxundefined [1]{%
 \@ifx{#1\undefined}
}%
\providecommand \@ifnum [1]{%
 \ifnum #1\expandafter \@firstoftwo
 \else \expandafter \@secondoftwo
 \fi
}%
\providecommand \@ifx [1]{%
 \ifx #1\expandafter \@firstoftwo
 \else \expandafter \@secondoftwo
 \fi
}%
\providecommand \natexlab [1]{#1}%
\providecommand \enquote  [1]{``#1''}%
\providecommand \bibnamefont  [1]{#1}%
\providecommand \bibfnamefont [1]{#1}%
\providecommand \citenamefont [1]{#1}%
\providecommand \href@noop [0]{\@secondoftwo}%
\providecommand \href [0]{\begingroup \@sanitize@url \@href}%
\providecommand \@href[1]{\@@startlink{#1}\@@href}%
\providecommand \@@href[1]{\endgroup#1\@@endlink}%
\providecommand \@sanitize@url [0]{\catcode `\\12\catcode `\$12\catcode
  `\&12\catcode `\#12\catcode `\^12\catcode `\_12\catcode `\%12\relax}%
\providecommand \@@startlink[1]{}%
\providecommand \@@endlink[0]{}%
\providecommand \url  [0]{\begingroup\@sanitize@url \@url }%
\providecommand \@url [1]{\endgroup\@href {#1}{\urlprefix }}%
\providecommand \urlprefix  [0]{URL }%
\providecommand \Eprint [0]{\href }%
\providecommand \doibase [0]{https://doi.org/}%
\providecommand \selectlanguage [0]{\@gobble}%
\providecommand \bibinfo  [0]{\@secondoftwo}%
\providecommand \bibfield  [0]{\@secondoftwo}%
\providecommand \translation [1]{[#1]}%
\providecommand \BibitemOpen [0]{}%
\providecommand \bibitemStop [0]{}%
\providecommand \bibitemNoStop [0]{.\EOS\space}%
\providecommand \EOS [0]{\spacefactor3000\relax}%
\providecommand \BibitemShut  [1]{\csname bibitem#1\endcsname}%
\let\auto@bib@innerbib\@empty
\bibitem [{\citenamefont {Deutsch}(2020)}]{PRXQuantum.1.020101}%
  \BibitemOpen
  \bibfield  {author} {\bibinfo {author} {\bibfnamefont {I.~H.}\ \bibnamefont
  {Deutsch}},\ }\href {https://doi.org/10.1103/PRXQuantum.1.020101} {\bibfield
  {journal} {\bibinfo  {journal} {PRX Quantum}\ }\textbf {\bibinfo {volume}
  {1}},\ \bibinfo {pages} {020101} (\bibinfo {year} {2020})}\BibitemShut
  {NoStop}%
\bibitem [{\citenamefont {Bell}()}]{Bell:1987hh}%
  \BibitemOpen
  \bibfield  {author} {\bibinfo {author} {\bibfnamefont {J.}~\bibnamefont
  {Bell}},\ }\href@noop {} {\emph {\bibinfo {title} {{Speakable and Unspeakable
  in Quantum Mechanics}}}}\ (\bibinfo  {publisher} {Cambridge University
  Press})\BibitemShut {NoStop}%
\bibitem [{\citenamefont {Aharonov}\ and\ \citenamefont
  {Rohrlich}()}]{Aharonov:2005}%
  \BibitemOpen
  \bibfield  {author} {\bibinfo {author} {\bibfnamefont {Y.}~\bibnamefont
  {Aharonov}}\ and\ \bibinfo {author} {\bibfnamefont {D.}~\bibnamefont
  {Rohrlich}},\ }\href@noop {} {\emph {\bibinfo {title} {{Quantum Paradoxes:
  Quantum Theory for the Perplexed}}}}\ (\bibinfo  {publisher}
  {Wiley-VCH})\BibitemShut {NoStop}%
\bibitem [{\citenamefont {'t~Hooft}(2014)}]{Hooft:2014kka}%
  \BibitemOpen
  \bibfield  {author} {\bibinfo {author} {\bibfnamefont {G.}~\bibnamefont
  {'t~Hooft}},\ }\href@noop {} {\  (\bibinfo {year} {2014})},\ \Eprint
  {https://arxiv.org/abs/1405.1548} {arXiv:1405.1548 [quant-ph]} \BibitemShut
  {NoStop}%
\bibitem [{\citenamefont {Penrose}(2014)}]{Penrose:2014nha}%
  \BibitemOpen
  \bibfield  {author} {\bibinfo {author} {\bibfnamefont {R.}~\bibnamefont
  {Penrose}},\ }\href {https://doi.org/10.1007/s10701-013-9770-0} {\bibfield
  {journal} {\bibinfo  {journal} {Found. Phys.}\ }\textbf {\bibinfo {volume}
  {44}},\ \bibinfo {pages} {557} (\bibinfo {year} {2014})}\BibitemShut
  {NoStop}%
\bibitem [{\citenamefont {Gell-Mann}\ and\ \citenamefont
  {Hartle}(2014)}]{Gell-Mann:2013hza}%
  \BibitemOpen
  \bibfield  {author} {\bibinfo {author} {\bibfnamefont {M.}~\bibnamefont
  {Gell-Mann}}\ and\ \bibinfo {author} {\bibfnamefont {J.~B.}\ \bibnamefont
  {Hartle}},\ }\href {https://doi.org/10.1103/PhysRevA.89.052125} {\bibfield
  {journal} {\bibinfo  {journal} {Phys. Rev. A}\ }\textbf {\bibinfo {volume}
  {89}},\ \bibinfo {pages} {052125} (\bibinfo {year} {2014})},\ \Eprint
  {https://arxiv.org/abs/1312.7454} {arXiv:1312.7454 [quant-ph]} \BibitemShut
  {NoStop}%
\bibitem [{\citenamefont {Weinberg}(2016)}]{Weinberg:2016axv}%
  \BibitemOpen
  \bibfield  {author} {\bibinfo {author} {\bibfnamefont {S.}~\bibnamefont
  {Weinberg}},\ }\href {https://doi.org/10.1103/PhysRevA.93.032124} {\bibfield
  {journal} {\bibinfo  {journal} {Phys. Rev. A}\ }\textbf {\bibinfo {volume}
  {93}},\ \bibinfo {pages} {032124} (\bibinfo {year} {2016})},\ \Eprint
  {https://arxiv.org/abs/1603.06008} {arXiv:1603.06008 [quant-ph]} \BibitemShut
  {NoStop}%
\bibitem [{\citenamefont {Leggett}(2007)}]{Leggett:2007zz}%
  \BibitemOpen
  \bibfield  {author} {\bibinfo {author} {\bibfnamefont {A.}~\bibnamefont
  {Leggett}},\ }\href {https://doi.org/10.1143/PTPS.170.100} {\bibfield
  {journal} {\bibinfo  {journal} {Prog. Theor. Phys. Suppl.}\ }\textbf
  {\bibinfo {volume} {170}},\ \bibinfo {pages} {100} (\bibinfo {year}
  {2007})}\BibitemShut {NoStop}%
\bibitem [{\citenamefont {Raghavan}\ \emph {et~al.}(2012)\citenamefont
  {Raghavan}, \citenamefont {Minic}, \citenamefont {Takeuchi},\ and\
  \citenamefont {Tze}}]{Raghavan:2012sy}%
  \BibitemOpen
  \bibfield  {author} {\bibinfo {author} {\bibfnamefont {R.~S.}\ \bibnamefont
  {Raghavan}}, \bibinfo {author} {\bibfnamefont {D.}~\bibnamefont {Minic}},
  \bibinfo {author} {\bibfnamefont {T.}~\bibnamefont {Takeuchi}},\ and\
  \bibinfo {author} {\bibfnamefont {C.~H.}\ \bibnamefont {Tze}},\ }\href@noop
  {} {\  (\bibinfo {year} {2012})},\ \Eprint {https://arxiv.org/abs/1210.5639}
  {arXiv:1210.5639 [hep-ph]} \BibitemShut {NoStop}%
\bibitem [{\citenamefont {Stueckelberg}(1960)}]{Stueckelberg:1960}%
  \BibitemOpen
  \bibfield  {author} {\bibinfo {author} {\bibfnamefont {E.~C.~G.}\
  \bibnamefont {Stueckelberg}},\ }\href@noop {} {\bibfield  {journal} {\bibinfo
   {journal} {Helv. Phys. Acta}\ }\textbf {\bibinfo {volume} {33}},\ \bibinfo
  {pages} {727} (\bibinfo {year} {1960})}\BibitemShut {NoStop}%
\bibitem [{\citenamefont {Adler}(1995)}]{Adler:1988hb}%
  \BibitemOpen
  \bibfield  {author} {\bibinfo {author} {\bibfnamefont {S.~L.}\ \bibnamefont
  {Adler}},\ }\href@noop {} {\emph {\bibinfo {title} {{Quaternionic quantum
  mechanics and quantum fields}}}}\ (\bibinfo  {publisher} {Oxford University
  Press},\ \bibinfo {address} {Oxford, UK},\ \bibinfo {year}
  {1995})\BibitemShut {NoStop}%
\bibitem [{\citenamefont {Gunaydin}\ \emph {et~al.}(1978)\citenamefont
  {Gunaydin}, \citenamefont {Piron},\ and\ \citenamefont
  {Ruegg}}]{Gunaydin:1978jq}%
  \BibitemOpen
  \bibfield  {author} {\bibinfo {author} {\bibfnamefont {M.}~\bibnamefont
  {Gunaydin}}, \bibinfo {author} {\bibfnamefont {C.}~\bibnamefont {Piron}},\
  and\ \bibinfo {author} {\bibfnamefont {H.}~\bibnamefont {Ruegg}},\ }\href
  {https://doi.org/10.1007/BF01609468} {\bibfield  {journal} {\bibinfo
  {journal} {Commun. Math. Phys.}\ }\textbf {\bibinfo {volume} {61}},\ \bibinfo
  {pages} {69} (\bibinfo {year} {1978})}\BibitemShut {NoStop}%
\bibitem [{\citenamefont {Okubo}(2011)}]{Okubo:1990nv}%
  \BibitemOpen
  \bibfield  {author} {\bibinfo {author} {\bibfnamefont {S.}~\bibnamefont
  {Okubo}},\ }\href@noop {} {\emph {\bibinfo {title} {{Introduction to Octonion
  and other Non-Associative Algebras in Physics}}}},\ Montroll Memorial Lecture
  Series in Mathematical Physics\ (\bibinfo  {publisher} {Cambridge University
  Press},\ \bibinfo {address} {Cambridge, UK},\ \bibinfo {year}
  {2011})\BibitemShut {NoStop}%
\bibitem [{\citenamefont {Gursey}\ and\ \citenamefont
  {Tze}(1996)}]{Gursey:1996mj}%
  \BibitemOpen
  \bibfield  {author} {\bibinfo {author} {\bibfnamefont {F.}~\bibnamefont
  {Gursey}}\ and\ \bibinfo {author} {\bibfnamefont {C.~H.}\ \bibnamefont
  {Tze}},\ }\href@noop {} {\emph {\bibinfo {title} {{On the role of division,
  Jordan and related algebras in particle physics}}}}\ (\bibinfo  {publisher}
  {World Scientific},\ \bibinfo {address} {Singapore},\ \bibinfo {year}
  {1996})\BibitemShut {NoStop}%
\bibitem [{\citenamefont {Chang}\ \emph
  {et~al.}(2013{\natexlab{a}})\citenamefont {Chang}, \citenamefont {Lewis},
  \citenamefont {Minic},\ and\ \citenamefont {Takeuchi}}]{Chang:2012eh}%
  \BibitemOpen
  \bibfield  {author} {\bibinfo {author} {\bibfnamefont {L.~N.}\ \bibnamefont
  {Chang}}, \bibinfo {author} {\bibfnamefont {Z.}~\bibnamefont {Lewis}},
  \bibinfo {author} {\bibfnamefont {D.}~\bibnamefont {Minic}},\ and\ \bibinfo
  {author} {\bibfnamefont {T.}~\bibnamefont {Takeuchi}},\ }\href
  {https://doi.org/10.1142/S0217984913500644} {\bibfield  {journal} {\bibinfo
  {journal} {Mod. Phys. Lett. B}\ }\textbf {\bibinfo {volume} {27}},\ \bibinfo
  {pages} {1350064} (\bibinfo {year} {2013}{\natexlab{a}})},\ \Eprint
  {https://arxiv.org/abs/1205.4800} {arXiv:1205.4800 [quant-ph]} \BibitemShut
  {NoStop}%
\bibitem [{\citenamefont {Chang}\ \emph
  {et~al.}(2013{\natexlab{b}})\citenamefont {Chang}, \citenamefont {Lewis},
  \citenamefont {Minic},\ and\ \citenamefont {Takeuchi}}]{Chang:2012gg}%
  \BibitemOpen
  \bibfield  {author} {\bibinfo {author} {\bibfnamefont {L.~N.}\ \bibnamefont
  {Chang}}, \bibinfo {author} {\bibfnamefont {Z.}~\bibnamefont {Lewis}},
  \bibinfo {author} {\bibfnamefont {D.}~\bibnamefont {Minic}},\ and\ \bibinfo
  {author} {\bibfnamefont {T.}~\bibnamefont {Takeuchi}},\ }\href
  {https://doi.org/10.1088/1751-8113/46/6/065304} {\bibfield  {journal}
  {\bibinfo  {journal} {J. Phys. A}\ }\textbf {\bibinfo {volume} {46}},\
  \bibinfo {pages} {065304} (\bibinfo {year} {2013}{\natexlab{b}})},\ \Eprint
  {https://arxiv.org/abs/1206.0064} {arXiv:1206.0064 [quant-ph]} \BibitemShut
  {NoStop}%
\bibitem [{\citenamefont {Chang}\ \emph
  {et~al.}(2013{\natexlab{c}})\citenamefont {Chang}, \citenamefont {Lewis},
  \citenamefont {Minic},\ and\ \citenamefont {Takeuchi}}]{Chang:2012we}%
  \BibitemOpen
  \bibfield  {author} {\bibinfo {author} {\bibfnamefont {L.~N.}\ \bibnamefont
  {Chang}}, \bibinfo {author} {\bibfnamefont {Z.}~\bibnamefont {Lewis}},
  \bibinfo {author} {\bibfnamefont {D.}~\bibnamefont {Minic}},\ and\ \bibinfo
  {author} {\bibfnamefont {T.}~\bibnamefont {Takeuchi}},\ }\href
  {https://doi.org/10.1088/1751-8113/46/48/485306} {\bibfield  {journal}
  {\bibinfo  {journal} {J. Phys. A}\ }\textbf {\bibinfo {volume} {46}},\
  \bibinfo {pages} {485306} (\bibinfo {year} {2013}{\natexlab{c}})},\ \Eprint
  {https://arxiv.org/abs/1208.5189} {arXiv:1208.5189 [math-ph]} \BibitemShut
  {NoStop}%
\bibitem [{\citenamefont {Takeuchi}\ \emph {et~al.}(2012)\citenamefont
  {Takeuchi}, \citenamefont {Chang}, \citenamefont {Lewis},\ and\ \citenamefont
  {Minic}}]{Takeuchi:2012mra}%
  \BibitemOpen
  \bibfield  {author} {\bibinfo {author} {\bibfnamefont {T.}~\bibnamefont
  {Takeuchi}}, \bibinfo {author} {\bibfnamefont {L.~N.}\ \bibnamefont {Chang}},
  \bibinfo {author} {\bibfnamefont {Z.}~\bibnamefont {Lewis}},\ and\ \bibinfo
  {author} {\bibfnamefont {D.}~\bibnamefont {Minic}},\ }\href
  {https://doi.org/10.1063/1.4773173} {\bibfield  {journal} {\bibinfo
  {journal} {AIP Conf. Proc.}\ }\textbf {\bibinfo {volume} {1508}},\ \bibinfo
  {pages} {502} (\bibinfo {year} {2012})},\ \Eprint
  {https://arxiv.org/abs/1208.5544} {arXiv:1208.5544 [quant-ph]} \BibitemShut
  {NoStop}%
\bibitem [{\citenamefont {Chang}\ \emph
  {et~al.}(2014{\natexlab{a}})\citenamefont {Chang}, \citenamefont {Lewis},
  \citenamefont {Minic},\ and\ \citenamefont {Takeuchi}}]{Chang:2013joa}%
  \BibitemOpen
  \bibfield  {author} {\bibinfo {author} {\bibfnamefont {L.~N.}\ \bibnamefont
  {Chang}}, \bibinfo {author} {\bibfnamefont {Z.}~\bibnamefont {Lewis}},
  \bibinfo {author} {\bibfnamefont {D.}~\bibnamefont {Minic}},\ and\ \bibinfo
  {author} {\bibfnamefont {T.}~\bibnamefont {Takeuchi}},\ }\href
  {https://doi.org/10.1142/S0217751X14300063} {\bibfield  {journal} {\bibinfo
  {journal} {Int. J. Mod. Phys. A}\ }\textbf {\bibinfo {volume} {29}},\
  \bibinfo {pages} {1430006} (\bibinfo {year} {2014}{\natexlab{a}})},\ \Eprint
  {https://arxiv.org/abs/1312.0645} {arXiv:1312.0645 [quant-ph]} \BibitemShut
  {NoStop}%
\bibitem [{\citenamefont {Chang}\ \emph
  {et~al.}(2014{\natexlab{b}})\citenamefont {Chang}, \citenamefont {Lewis},
  \citenamefont {Minic},\ and\ \citenamefont {Takeuchi}}]{Chang:2013rya}%
  \BibitemOpen
  \bibfield  {author} {\bibinfo {author} {\bibfnamefont {L.~N.}\ \bibnamefont
  {Chang}}, \bibinfo {author} {\bibfnamefont {Z.}~\bibnamefont {Lewis}},
  \bibinfo {author} {\bibfnamefont {D.}~\bibnamefont {Minic}},\ and\ \bibinfo
  {author} {\bibfnamefont {T.}~\bibnamefont {Takeuchi}},\ }\href
  {https://doi.org/10.1088/1751-8113/47/40/405304} {\bibfield  {journal}
  {\bibinfo  {journal} {J. Phys. A}\ }\textbf {\bibinfo {volume} {47}},\
  \bibinfo {pages} {405304} (\bibinfo {year} {2014}{\natexlab{b}})},\ \Eprint
  {https://arxiv.org/abs/1312.4191} {arXiv:1312.4191 [quant-ph]} \BibitemShut
  {NoStop}%
\bibitem [{\citenamefont {Chang}\ \emph {et~al.}(2019)\citenamefont {Chang},
  \citenamefont {Minic},\ and\ \citenamefont {Takeuchi}}]{Chang:2019kcp}%
  \BibitemOpen
  \bibfield  {author} {\bibinfo {author} {\bibfnamefont {L.~N.}\ \bibnamefont
  {Chang}}, \bibinfo {author} {\bibfnamefont {D.}~\bibnamefont {Minic}},\ and\
  \bibinfo {author} {\bibfnamefont {T.}~\bibnamefont {Takeuchi}},\ }\href
  {https://doi.org/10.1088/1742-6596/1275/1/012036} {\bibfield  {journal}
  {\bibinfo  {journal} {J. Phys. Conf. Ser.}\ }\textbf {\bibinfo {volume}
  {1275}},\ \bibinfo {pages} {012036} (\bibinfo {year} {2019})},\ \Eprint
  {https://arxiv.org/abs/1903.06337} {arXiv:1903.06337 [quant-ph]} \BibitemShut
  {NoStop}%
\bibitem [{\citenamefont {Weinberg}(1989{\natexlab{a}})}]{Weinberg:1989cm}%
  \BibitemOpen
  \bibfield  {author} {\bibinfo {author} {\bibfnamefont {S.}~\bibnamefont
  {Weinberg}},\ }\href {https://doi.org/10.1103/PhysRevLett.62.485} {\bibfield
  {journal} {\bibinfo  {journal} {Phys. Rev. Lett.}\ }\textbf {\bibinfo
  {volume} {62}},\ \bibinfo {pages} {485} (\bibinfo {year}
  {1989}{\natexlab{a}})}\BibitemShut {NoStop}%
\bibitem [{\citenamefont {Weinberg}(1989{\natexlab{b}})}]{Weinberg:1989us}%
  \BibitemOpen
  \bibfield  {author} {\bibinfo {author} {\bibfnamefont {S.}~\bibnamefont
  {Weinberg}},\ }\href {https://doi.org/10.1016/0003-4916(89)90276-5}
  {\bibfield  {journal} {\bibinfo  {journal} {Annals Phys.}\ }\textbf {\bibinfo
  {volume} {194}},\ \bibinfo {pages} {336} (\bibinfo {year}
  {1989}{\natexlab{b}})}\BibitemShut {NoStop}%
\bibitem [{\citenamefont {Minic}\ and\ \citenamefont
  {Tze}(2002)}]{Minic:2002pd}%
  \BibitemOpen
  \bibfield  {author} {\bibinfo {author} {\bibfnamefont {D.}~\bibnamefont
  {Minic}}\ and\ \bibinfo {author} {\bibfnamefont {C.~H.}\ \bibnamefont
  {Tze}},\ }\href {https://doi.org/10.1016/S0370-2693(02)01865-8} {\bibfield
  {journal} {\bibinfo  {journal} {Phys. Lett. B}\ }\textbf {\bibinfo {volume}
  {536}},\ \bibinfo {pages} {305} (\bibinfo {year} {2002})},\ \Eprint
  {https://arxiv.org/abs/hep-th/0202173} {arXiv:hep-th/0202173} \BibitemShut
  {NoStop}%
\bibitem [{\citenamefont {Nambu}(1973)}]{Nambu:1973qe}%
  \BibitemOpen
  \bibfield  {author} {\bibinfo {author} {\bibfnamefont {Y.}~\bibnamefont
  {Nambu}},\ }\href {https://doi.org/10.1103/PhysRevD.7.2405} {\bibfield
  {journal} {\bibinfo  {journal} {Phys. Rev. D}\ }\textbf {\bibinfo {volume}
  {7}},\ \bibinfo {pages} {2405} (\bibinfo {year} {1973})}\BibitemShut
  {NoStop}%
\bibitem [{\citenamefont {Kibble}(1979)}]{Kibble:1978tm}%
  \BibitemOpen
  \bibfield  {author} {\bibinfo {author} {\bibfnamefont {T.}~\bibnamefont
  {Kibble}},\ }\href {https://doi.org/10.1007/BF01225149} {\bibfield  {journal}
  {\bibinfo  {journal} {Commun. Math. Phys.}\ }\textbf {\bibinfo {volume}
  {65}},\ \bibinfo {pages} {189} (\bibinfo {year} {1979})}\BibitemShut
  {NoStop}%
\bibitem [{Note1()}]{Note1}%
  \BibitemOpen
  \bibinfo {note} {Mathematica encodes $\protect \mathrm {sn}(u,k)$, $\protect
  \mathrm {cn}(u,k)$, and $\protect \mathrm {dn}(u,k)$ respectively as \protect
  \texttt {JacobiSN[$u$,$m$]}, \protect \texttt {JacobiCN[$u$,$m$]}, and
  \protect \texttt {JacobiDN[$u$,$m$]} with $m=k^2$.}\BibitemShut {Stop}%
\bibitem [{Note2()}]{Note2}%
  \BibitemOpen
  \bibinfo {note} {Mathematica encodes $K(k)$ as \protect \texttt
  {EllipticK[$m$]} with $m=k^2$.}\BibitemShut {Stop}%
\bibitem [{\citenamefont {Opatrn\'{y}}\ \emph {et~al.}(2018)\citenamefont
  {Opatrn\'{y}}, \citenamefont {Richterek},\ and\ \citenamefont
  {Opatrn\'{y}}}]{Opatrny:2018}%
  \BibitemOpen
  \bibfield  {author} {\bibinfo {author} {\bibfnamefont {T.}~\bibnamefont
  {Opatrn\'{y}}}, \bibinfo {author} {\bibfnamefont {L.}~\bibnamefont
  {Richterek}},\ and\ \bibinfo {author} {\bibfnamefont {M.}~\bibnamefont
  {Opatrn\'{y}}},\ }\href {https://doi.org/10.1038/s41598-018-20486-y}
  {\bibfield  {journal} {\bibinfo  {journal} {Scientific Reports}\ }\textbf
  {\bibinfo {volume} {8}},\ \bibinfo {pages} {1984} (\bibinfo {year} {2018})},\
  \Eprint {https://arxiv.org/abs/1708.07764} {arXiv:1708.07764 [quant-ph]}
  \BibitemShut {NoStop}%
\bibitem [{\citenamefont {Gradshteyn}\ and\ \citenamefont
  {Ryzhik}(2014)}]{Gradshteyn-Ryzhik}%
  \BibitemOpen
  \bibfield  {author} {\bibinfo {author} {\bibfnamefont {I.~D.}\ \bibnamefont
  {Gradshteyn}}\ and\ \bibinfo {author} {\bibfnamefont {I.~M.}\ \bibnamefont
  {Ryzhik}},\ }\href@noop {} {\emph {\bibinfo {title} {{Table of Integrals,
  Series, and Products}}}},\ \bibinfo {edition} {corrected and enlarged}\ ed.,\
  edited by\ \bibinfo {editor} {\bibfnamefont {A.}~\bibnamefont {Jeffery}}\
  (\bibinfo  {publisher} {Academic Press},\ \bibinfo {year} {2014})\BibitemShut
  {NoStop}%
\bibitem [{\citenamefont {Abramowitz}\ and\ \citenamefont
  {Stegun}(2014)}]{Abramowitz-Stegun}%
  \BibitemOpen
  \bibfield  {author} {\bibinfo {author} {\bibfnamefont {M.}~\bibnamefont
  {Abramowitz}}\ and\ \bibinfo {author} {\bibfnamefont {I.}~\bibnamefont
  {Stegun}},\ }\href@noop {} {\emph {\bibinfo {title} {{Handbook of
  Mathematical Functions with Formulas, Graphs, and Mathematical Tables}}}},\
  \bibinfo {edition} {illustrated}\ ed.\ (\bibinfo  {publisher} {Martino Fine
  Books},\ \bibinfo {year} {2014})\BibitemShut {NoStop}%
\bibitem [{\citenamefont {Whittaker}\ and\ \citenamefont
  {Watson}(2020)}]{Whittaker-Watson}%
  \BibitemOpen
  \bibfield  {author} {\bibinfo {author} {\bibfnamefont {E.~T.}\ \bibnamefont
  {Whittaker}}\ and\ \bibinfo {author} {\bibfnamefont {G.~N.}\ \bibnamefont
  {Watson}},\ }\href@noop {} {\emph {\bibinfo {title} {{A Course of Modern
  Analysis}}}},\ \bibinfo {edition} {3rd}\ ed.\ (\bibinfo  {publisher} {Dover
  Publications},\ \bibinfo {year} {2020})\BibitemShut {NoStop}%
\bibitem [{\citenamefont {Kayser}(2012)}]{Kayser:2011jn}%
  \BibitemOpen
  \bibfield  {author} {\bibinfo {author} {\bibfnamefont {B.}~\bibnamefont
  {Kayser}},\ }\href {https://doi.org/10.1063/1.3700586} {\bibfield  {journal}
  {\bibinfo  {journal} {AIP Conf. Proc.}\ }\textbf {\bibinfo {volume} {1441}},\
  \bibinfo {pages} {464} (\bibinfo {year} {2012})},\ \Eprint
  {https://arxiv.org/abs/1110.3047} {arXiv:1110.3047 [hep-ph]} \BibitemShut
  {NoStop}%
\bibitem [{\citenamefont {de~Salas}\ \emph {et~al.}(2020)\citenamefont
  {de~Salas}, \citenamefont {Forero}, \citenamefont {Gariazzo}, \citenamefont
  {Mart\'\i{}nez-Mirav\'e}, \citenamefont {Mena}, \citenamefont {Ternes},
  \citenamefont {T\'ortola},\ and\ \citenamefont {Valle}}]{deSalas:2020pgw}%
  \BibitemOpen
  \bibfield  {author} {\bibinfo {author} {\bibfnamefont {P.}~\bibnamefont
  {de~Salas}}, \bibinfo {author} {\bibfnamefont {D.}~\bibnamefont {Forero}},
  \bibinfo {author} {\bibfnamefont {S.}~\bibnamefont {Gariazzo}}, \bibinfo
  {author} {\bibfnamefont {P.}~\bibnamefont {Mart\'\i{}nez-Mirav\'e}}, \bibinfo
  {author} {\bibfnamefont {O.}~\bibnamefont {Mena}}, \bibinfo {author}
  {\bibfnamefont {C.}~\bibnamefont {Ternes}}, \bibinfo {author} {\bibfnamefont
  {M.}~\bibnamefont {T\'ortola}},\ and\ \bibinfo {author} {\bibfnamefont
  {J.}~\bibnamefont {Valle}},\ }\href@noop {} {\  (\bibinfo {year} {2020})},\
  \Eprint {https://arxiv.org/abs/2006.11237} {arXiv:2006.11237 [hep-ph]}
  \BibitemShut {NoStop}%
\bibitem [{\citenamefont {Terliuk}(2019)}]{Terliuk:2019hkb}%
  \BibitemOpen
  \bibfield  {author} {\bibinfo {author} {\bibfnamefont {A.}~\bibnamefont
  {Terliuk}} (\bibinfo {collaboration} {IceCube}),\ }\href
  {https://doi.org/10.22323/1.337.0007} {\bibfield  {journal} {\bibinfo
  {journal} {PoS}\ }\textbf {\bibinfo {volume} {NOW2018}},\ \bibinfo {pages}
  {007} (\bibinfo {year} {2019})}\BibitemShut {NoStop}%
\bibitem [{\citenamefont {Guo}(2017)}]{Guo:2017mvv}%
  \BibitemOpen
  \bibfield  {author} {\bibinfo {author} {\bibfnamefont {W.-l.}\ \bibnamefont
  {Guo}} (\bibinfo {collaboration} {JUNO}),\ }\href
  {https://doi.org/10.1088/1742-6596/888/1/012205} {\bibfield  {journal}
  {\bibinfo  {journal} {J. Phys. Conf. Ser.}\ }\textbf {\bibinfo {volume}
  {888}},\ \bibinfo {pages} {012205} (\bibinfo {year} {2017})}\BibitemShut
  {NoStop}%
\bibitem [{\citenamefont {Higuera}(2018)}]{Higuera:2018zjz}%
  \BibitemOpen
  \bibfield  {author} {\bibinfo {author} {\bibfnamefont {A.}~\bibnamefont
  {Higuera}} (\bibinfo {collaboration} {DUNE}),\ }\href
  {https://doi.org/10.22323/1.314.0115} {\bibfield  {journal} {\bibinfo
  {journal} {PoS}\ }\textbf {\bibinfo {volume} {EPS-HEP2017}},\ \bibinfo
  {pages} {115} (\bibinfo {year} {2018})}\BibitemShut {NoStop}%
\bibitem [{\citenamefont {Takhtajan}(1994)}]{Takhtajan:1993vr}%
  \BibitemOpen
  \bibfield  {author} {\bibinfo {author} {\bibfnamefont {L.}~\bibnamefont
  {Takhtajan}},\ }\href {https://doi.org/10.1007/BF02103278} {\bibfield
  {journal} {\bibinfo  {journal} {Commun. Math. Phys.}\ }\textbf {\bibinfo
  {volume} {160}},\ \bibinfo {pages} {295} (\bibinfo {year} {1994})},\ \Eprint
  {https://arxiv.org/abs/hep-th/9301111} {arXiv:hep-th/9301111} \BibitemShut
  {NoStop}%
\bibitem [{\citenamefont {Dito}\ \emph {et~al.}(1997)\citenamefont {Dito},
  \citenamefont {Flato}, \citenamefont {Sternheimer},\ and\ \citenamefont
  {Takhtajan}}]{Dito:1996xr}%
  \BibitemOpen
  \bibfield  {author} {\bibinfo {author} {\bibfnamefont {G.}~\bibnamefont
  {Dito}}, \bibinfo {author} {\bibfnamefont {M.}~\bibnamefont {Flato}},
  \bibinfo {author} {\bibfnamefont {D.}~\bibnamefont {Sternheimer}},\ and\
  \bibinfo {author} {\bibfnamefont {L.}~\bibnamefont {Takhtajan}},\ }\href
  {https://doi.org/10.1007/BF02509794} {\bibfield  {journal} {\bibinfo
  {journal} {Commun. Math. Phys.}\ }\textbf {\bibinfo {volume} {183}},\
  \bibinfo {pages} {1} (\bibinfo {year} {1997})},\ \Eprint
  {https://arxiv.org/abs/hep-th/9602016} {arXiv:hep-th/9602016} \BibitemShut
  {NoStop}%
\bibitem [{\citenamefont {Dito}\ and\ \citenamefont
  {Flato}(1997)}]{Dito:1996hn}%
  \BibitemOpen
  \bibfield  {author} {\bibinfo {author} {\bibfnamefont {G.}~\bibnamefont
  {Dito}}\ and\ \bibinfo {author} {\bibfnamefont {M.}~\bibnamefont {Flato}},\
  }\href {https://doi.org/10.1023/A:1007309124218} {\bibfield  {journal}
  {\bibinfo  {journal} {Lett. Math. Phys.}\ }\textbf {\bibinfo {volume} {39}},\
  \bibinfo {pages} {107} (\bibinfo {year} {1997})},\ \Eprint
  {https://arxiv.org/abs/hep-th/9609114} {arXiv:hep-th/9609114} \BibitemShut
  {NoStop}%
\bibitem [{\citenamefont {Bergshoeff}\ \emph {et~al.}(1990)\citenamefont
  {Bergshoeff}, \citenamefont {Sezgin}, \citenamefont {Tanii},\ and\
  \citenamefont {Townsend}}]{Bergshoeff:1988hw}%
  \BibitemOpen
  \bibfield  {author} {\bibinfo {author} {\bibfnamefont {E.}~\bibnamefont
  {Bergshoeff}}, \bibinfo {author} {\bibfnamefont {E.}~\bibnamefont {Sezgin}},
  \bibinfo {author} {\bibfnamefont {Y.}~\bibnamefont {Tanii}},\ and\ \bibinfo
  {author} {\bibfnamefont {P.}~\bibnamefont {Townsend}},\ }\href
  {https://doi.org/10.1016/0003-4916(90)90381-W} {\bibfield  {journal}
  {\bibinfo  {journal} {Annals Phys.}\ }\textbf {\bibinfo {volume} {199}},\
  \bibinfo {pages} {340} (\bibinfo {year} {1990})}\BibitemShut {NoStop}%
\bibitem [{\citenamefont {Awata}\ and\ \citenamefont
  {Minic}(1998)}]{Awata:1997yr}%
  \BibitemOpen
  \bibfield  {author} {\bibinfo {author} {\bibfnamefont {H.}~\bibnamefont
  {Awata}}\ and\ \bibinfo {author} {\bibfnamefont {D.}~\bibnamefont {Minic}},\
  }\href {https://doi.org/10.1088/1126-6708/1998/04/006} {\bibfield  {journal}
  {\bibinfo  {journal} {JHEP}\ }\textbf {\bibinfo {volume} {04}},\ \bibinfo
  {pages} {006}},\ \Eprint {https://arxiv.org/abs/hep-th/9711034}
  {arXiv:hep-th/9711034} \BibitemShut {NoStop}%
\bibitem [{\citenamefont {Awata}\ \emph {et~al.}(2001)\citenamefont {Awata},
  \citenamefont {Li}, \citenamefont {Minic},\ and\ \citenamefont
  {Yoneya}}]{Awata:1999dz}%
  \BibitemOpen
  \bibfield  {author} {\bibinfo {author} {\bibfnamefont {H.}~\bibnamefont
  {Awata}}, \bibinfo {author} {\bibfnamefont {M.}~\bibnamefont {Li}}, \bibinfo
  {author} {\bibfnamefont {D.}~\bibnamefont {Minic}},\ and\ \bibinfo {author}
  {\bibfnamefont {T.}~\bibnamefont {Yoneya}},\ }\href
  {https://doi.org/10.1088/1126-6708/2001/02/013} {\bibfield  {journal}
  {\bibinfo  {journal} {JHEP}\ }\textbf {\bibinfo {volume} {02}},\ \bibinfo
  {pages} {013}},\ \Eprint {https://arxiv.org/abs/hep-th/9906248}
  {arXiv:hep-th/9906248} \BibitemShut {NoStop}%
\bibitem [{\citenamefont {Fujikawa}\ and\ \citenamefont
  {Okuyama}(1997)}]{Fujikawa:1997jt}%
  \BibitemOpen
  \bibfield  {author} {\bibinfo {author} {\bibfnamefont {K.}~\bibnamefont
  {Fujikawa}}\ and\ \bibinfo {author} {\bibfnamefont {K.}~\bibnamefont
  {Okuyama}},\ }\href {https://doi.org/10.1016/S0370-2693(97)01027-7}
  {\bibfield  {journal} {\bibinfo  {journal} {Phys. Lett. B}\ }\textbf
  {\bibinfo {volume} {411}},\ \bibinfo {pages} {261} (\bibinfo {year}
  {1997})},\ \Eprint {https://arxiv.org/abs/hep-th/9706027}
  {arXiv:hep-th/9706027} \BibitemShut {NoStop}%
\bibitem [{\citenamefont {Smolin}(1998)}]{Smolin:1997ai}%
  \BibitemOpen
  \bibfield  {author} {\bibinfo {author} {\bibfnamefont {L.}~\bibnamefont
  {Smolin}},\ }\href {https://doi.org/10.1103/PhysRevD.57.6216} {\bibfield
  {journal} {\bibinfo  {journal} {Phys. Rev. D}\ }\textbf {\bibinfo {volume}
  {57}},\ \bibinfo {pages} {6216} (\bibinfo {year} {1998})},\ \Eprint
  {https://arxiv.org/abs/hep-th/9710191} {arXiv:hep-th/9710191} \BibitemShut
  {NoStop}%
\bibitem [{\citenamefont {Curtright}\ and\ \citenamefont
  {Zachos}(2003{\natexlab{a}})}]{Curtright:2002fd}%
  \BibitemOpen
  \bibfield  {author} {\bibinfo {author} {\bibfnamefont {T.~L.}\ \bibnamefont
  {Curtright}}\ and\ \bibinfo {author} {\bibfnamefont {C.~K.}\ \bibnamefont
  {Zachos}},\ }\href {https://doi.org/10.1103/PhysRevD.68.085001} {\bibfield
  {journal} {\bibinfo  {journal} {Phys. Rev. D}\ }\textbf {\bibinfo {volume}
  {68}},\ \bibinfo {pages} {085001} (\bibinfo {year} {2003}{\natexlab{a}})},\
  \Eprint {https://arxiv.org/abs/hep-th/0212267} {arXiv:hep-th/0212267}
  \BibitemShut {NoStop}%
\bibitem [{\citenamefont {Curtright}\ and\ \citenamefont
  {Zachos}(2002)}]{Curtright:2002sr}%
  \BibitemOpen
  \bibfield  {author} {\bibinfo {author} {\bibfnamefont {T.~L.}\ \bibnamefont
  {Curtright}}\ and\ \bibinfo {author} {\bibfnamefont {C.~K.}\ \bibnamefont
  {Zachos}},\ }\href {https://doi.org/10.1088/1367-2630/4/1/383} {\bibfield
  {journal} {\bibinfo  {journal} {New J. Phys.}\ }\textbf {\bibinfo {volume}
  {4}},\ \bibinfo {pages} {83} (\bibinfo {year} {2002})},\ \Eprint
  {https://arxiv.org/abs/hep-th/0205063} {arXiv:hep-th/0205063} \BibitemShut
  {NoStop}%
\bibitem [{\citenamefont {Curtright}\ and\ \citenamefont
  {Zachos}(2003{\natexlab{b}})}]{Curtright:2003fn}%
  \BibitemOpen
  \bibfield  {author} {\bibinfo {author} {\bibfnamefont {T.~L.}\ \bibnamefont
  {Curtright}}\ and\ \bibinfo {author} {\bibfnamefont {C.~K.}\ \bibnamefont
  {Zachos}},\ }\href {https://doi.org/10.1063/1.1594404} {\bibfield  {journal}
  {\bibinfo  {journal} {AIP Conf. Proc.}\ }\textbf {\bibinfo {volume} {672}},\
  \bibinfo {pages} {165} (\bibinfo {year} {2003}{\natexlab{b}})},\ \Eprint
  {https://arxiv.org/abs/hep-th/0303088} {arXiv:hep-th/0303088} \BibitemShut
  {NoStop}%
\bibitem [{\citenamefont {Bagger}\ and\ \citenamefont
  {Lambert}(2008)}]{Bagger:2007jr}%
  \BibitemOpen
  \bibfield  {author} {\bibinfo {author} {\bibfnamefont {J.}~\bibnamefont
  {Bagger}}\ and\ \bibinfo {author} {\bibfnamefont {N.}~\bibnamefont
  {Lambert}},\ }\href {https://doi.org/10.1103/PhysRevD.77.065008} {\bibfield
  {journal} {\bibinfo  {journal} {Phys. Rev. D}\ }\textbf {\bibinfo {volume}
  {77}},\ \bibinfo {pages} {065008} (\bibinfo {year} {2008})},\ \Eprint
  {https://arxiv.org/abs/0711.0955} {arXiv:0711.0955 [hep-th]} \BibitemShut
  {NoStop}%
\bibitem [{\citenamefont {Gustavsson}(2009)}]{Gustavsson:2007vu}%
  \BibitemOpen
  \bibfield  {author} {\bibinfo {author} {\bibfnamefont {A.}~\bibnamefont
  {Gustavsson}},\ }\href {https://doi.org/10.1016/j.nuclphysb.2008.11.014}
  {\bibfield  {journal} {\bibinfo  {journal} {Nucl. Phys. B}\ }\textbf
  {\bibinfo {volume} {811}},\ \bibinfo {pages} {66} (\bibinfo {year} {2009})},\
  \Eprint {https://arxiv.org/abs/0709.1260} {arXiv:0709.1260 [hep-th]}
  \BibitemShut {NoStop}%
\bibitem [{\citenamefont {Chu}\ \emph {et~al.}(2008)\citenamefont {Chu},
  \citenamefont {Ho}, \citenamefont {Matsuo},\ and\ \citenamefont
  {Shiba}}]{Chu:2008qv}%
  \BibitemOpen
  \bibfield  {author} {\bibinfo {author} {\bibfnamefont {C.-S.}\ \bibnamefont
  {Chu}}, \bibinfo {author} {\bibfnamefont {P.-M.}\ \bibnamefont {Ho}},
  \bibinfo {author} {\bibfnamefont {Y.}~\bibnamefont {Matsuo}},\ and\ \bibinfo
  {author} {\bibfnamefont {S.}~\bibnamefont {Shiba}},\ }\href
  {https://doi.org/10.1088/1126-6708/2008/08/076} {\bibfield  {journal}
  {\bibinfo  {journal} {JHEP}\ }\textbf {\bibinfo {volume} {08}},\ \bibinfo
  {pages} {076}},\ \Eprint {https://arxiv.org/abs/0807.0812} {arXiv:0807.0812
  [hep-th]} \BibitemShut {NoStop}%
\bibitem [{\citenamefont {Ho}\ and\ \citenamefont {Matsuo}(2016)}]{Ho:2016hob}%
  \BibitemOpen
  \bibfield  {author} {\bibinfo {author} {\bibfnamefont {P.-M.}\ \bibnamefont
  {Ho}}\ and\ \bibinfo {author} {\bibfnamefont {Y.}~\bibnamefont {Matsuo}},\
  }\href {https://doi.org/10.1093/ptep/ptw075} {\bibfield  {journal} {\bibinfo
  {journal} {PTEP}\ }\textbf {\bibinfo {volume} {2016}},\ \bibinfo {pages}
  {06A104} (\bibinfo {year} {2016})},\ \Eprint
  {https://arxiv.org/abs/1603.09534} {arXiv:1603.09534 [hep-th]} \BibitemShut
  {NoStop}%
\bibitem [{\citenamefont {Freidel}\ \emph
  {et~al.}(2019{\natexlab{a}})\citenamefont {Freidel}, \citenamefont {Leigh},\
  and\ \citenamefont {Minic}}]{Freidel:2019jor}%
  \BibitemOpen
  \bibfield  {author} {\bibinfo {author} {\bibfnamefont {L.}~\bibnamefont
  {Freidel}}, \bibinfo {author} {\bibfnamefont {R.~G.}\ \bibnamefont {Leigh}},\
  and\ \bibinfo {author} {\bibfnamefont {D.}~\bibnamefont {Minic}},\ }\href
  {https://doi.org/10.1142/9789811213144_0004} {\bibfield  {journal} {\bibinfo
  {journal} {Int. J. Mod. Phys. A}\ }\textbf {\bibinfo {volume} {34}},\
  \bibinfo {pages} {1941004} (\bibinfo {year}
  {2019}{\natexlab{a}})}\BibitemShut {NoStop}%
\bibitem [{\citenamefont {Minic}(2020)}]{Minic:2020oho}%
  \BibitemOpen
  \bibfield  {author} {\bibinfo {author} {\bibfnamefont {D.}~\bibnamefont
  {Minic}},\ }in\ \href@noop {} {\emph {\bibinfo {booktitle} {{10th
  MATHEMATICAL PHYSICS MEETING}: {School and Conference on Modern Mathematical
  Physics}}}}\ (\bibinfo {year} {2020})\ pp.\ \bibinfo {pages} {183--218},\
  \Eprint {https://arxiv.org/abs/2003.00318} {arXiv:2003.00318 [hep-th]}
  \BibitemShut {NoStop}%
\bibitem [{\citenamefont {Berglund}\ \emph {et~al.}(2019)\citenamefont
  {Berglund}, \citenamefont {H\"ubsch},\ and\ \citenamefont
  {Mini\'c}}]{Berglund:2019ctg}%
  \BibitemOpen
  \bibfield  {author} {\bibinfo {author} {\bibfnamefont {P.}~\bibnamefont
  {Berglund}}, \bibinfo {author} {\bibfnamefont {T.}~\bibnamefont {H\"ubsch}},\
  and\ \bibinfo {author} {\bibfnamefont {D.}~\bibnamefont {Mini\'c}},\ }\href
  {https://doi.org/10.1016/j.physletb.2019.134950} {\bibfield  {journal}
  {\bibinfo  {journal} {Phys. Lett. B}\ }\textbf {\bibinfo {volume} {798}},\
  \bibinfo {pages} {134950} (\bibinfo {year} {2019})},\ \Eprint
  {https://arxiv.org/abs/1905.08269} {arXiv:1905.08269 [hep-th]} \BibitemShut
  {NoStop}%
\bibitem [{\citenamefont {Berglund}\ \emph {et~al.}(2020)\citenamefont
  {Berglund}, \citenamefont {H\"ubsch},\ and\ \citenamefont
  {Minic}}]{Berglund:2020qcu}%
  \BibitemOpen
  \bibfield  {author} {\bibinfo {author} {\bibfnamefont {P.}~\bibnamefont
  {Berglund}}, \bibinfo {author} {\bibfnamefont {T.}~\bibnamefont {H\"ubsch}},\
  and\ \bibinfo {author} {\bibfnamefont {D.}~\bibnamefont {Minic}},\
  }\href@noop {} {\  (\bibinfo {year} {2020})},\ \Eprint
  {https://arxiv.org/abs/2010.15610} {arXiv:2010.15610 [hep-th]} \BibitemShut
  {NoStop}%
\bibitem [{\citenamefont {Freidel}\ \emph
  {et~al.}(2014{\natexlab{a}})\citenamefont {Freidel}, \citenamefont {Leigh},\
  and\ \citenamefont {Minic}}]{Freidel:2013zga}%
  \BibitemOpen
  \bibfield  {author} {\bibinfo {author} {\bibfnamefont {L.}~\bibnamefont
  {Freidel}}, \bibinfo {author} {\bibfnamefont {R.~G.}\ \bibnamefont {Leigh}},\
  and\ \bibinfo {author} {\bibfnamefont {D.}~\bibnamefont {Minic}},\ }\href
  {https://doi.org/10.1016/j.physletb.2014.01.067} {\bibfield  {journal}
  {\bibinfo  {journal} {Phys. Lett. B}\ }\textbf {\bibinfo {volume} {730}},\
  \bibinfo {pages} {302} (\bibinfo {year} {2014}{\natexlab{a}})},\ \Eprint
  {https://arxiv.org/abs/1307.7080} {arXiv:1307.7080 [hep-th]} \BibitemShut
  {NoStop}%
\bibitem [{\citenamefont {Freidel}\ \emph
  {et~al.}(2014{\natexlab{b}})\citenamefont {Freidel}, \citenamefont {Leigh},\
  and\ \citenamefont {Minic}}]{Freidel:2014qna}%
  \BibitemOpen
  \bibfield  {author} {\bibinfo {author} {\bibfnamefont {L.}~\bibnamefont
  {Freidel}}, \bibinfo {author} {\bibfnamefont {R.~G.}\ \bibnamefont {Leigh}},\
  and\ \bibinfo {author} {\bibfnamefont {D.}~\bibnamefont {Minic}},\ }\href
  {https://doi.org/10.1142/S0218271814420061} {\bibfield  {journal} {\bibinfo
  {journal} {Int. J. Mod. Phys. D}\ }\textbf {\bibinfo {volume} {23}},\
  \bibinfo {pages} {1442006} (\bibinfo {year} {2014}{\natexlab{b}})},\ \Eprint
  {https://arxiv.org/abs/1405.3949} {arXiv:1405.3949 [hep-th]} \BibitemShut
  {NoStop}%
\bibitem [{\citenamefont {Freidel}\ \emph {et~al.}(2015)\citenamefont
  {Freidel}, \citenamefont {Leigh},\ and\ \citenamefont
  {Minic}}]{Freidel:2015pka}%
  \BibitemOpen
  \bibfield  {author} {\bibinfo {author} {\bibfnamefont {L.}~\bibnamefont
  {Freidel}}, \bibinfo {author} {\bibfnamefont {R.~G.}\ \bibnamefont {Leigh}},\
  and\ \bibinfo {author} {\bibfnamefont {D.}~\bibnamefont {Minic}},\ }\href
  {https://doi.org/10.1007/JHEP06(2015)006} {\bibfield  {journal} {\bibinfo
  {journal} {JHEP}\ }\textbf {\bibinfo {volume} {06}},\ \bibinfo {pages}
  {006}},\ \Eprint {https://arxiv.org/abs/1502.08005} {arXiv:1502.08005
  [hep-th]} \BibitemShut {NoStop}%
\bibitem [{\citenamefont {Freidel}\ \emph {et~al.}(2016)\citenamefont
  {Freidel}, \citenamefont {Leigh},\ and\ \citenamefont
  {Minic}}]{Freidel:2016pls}%
  \BibitemOpen
  \bibfield  {author} {\bibinfo {author} {\bibfnamefont {L.}~\bibnamefont
  {Freidel}}, \bibinfo {author} {\bibfnamefont {R.~G.}\ \bibnamefont {Leigh}},\
  and\ \bibinfo {author} {\bibfnamefont {D.}~\bibnamefont {Minic}},\ }\href
  {https://doi.org/10.1103/PhysRevD.94.104052} {\bibfield  {journal} {\bibinfo
  {journal} {Phys. Rev. D}\ }\textbf {\bibinfo {volume} {94}},\ \bibinfo
  {pages} {104052} (\bibinfo {year} {2016})},\ \Eprint
  {https://arxiv.org/abs/1606.01829} {arXiv:1606.01829 [hep-th]} \BibitemShut
  {NoStop}%
\bibitem [{\citenamefont {Freidel}\ \emph
  {et~al.}(2017{\natexlab{a}})\citenamefont {Freidel}, \citenamefont {Leigh},\
  and\ \citenamefont {Minic}}]{Freidel:2017xsi}%
  \BibitemOpen
  \bibfield  {author} {\bibinfo {author} {\bibfnamefont {L.}~\bibnamefont
  {Freidel}}, \bibinfo {author} {\bibfnamefont {R.~G.}\ \bibnamefont {Leigh}},\
  and\ \bibinfo {author} {\bibfnamefont {D.}~\bibnamefont {Minic}},\ }\href
  {https://doi.org/10.1088/1742-6596/804/1/012032} {\bibfield  {journal}
  {\bibinfo  {journal} {J. Phys. Conf. Ser.}\ }\textbf {\bibinfo {volume}
  {804}},\ \bibinfo {pages} {012032} (\bibinfo {year}
  {2017}{\natexlab{a}})}\BibitemShut {NoStop}%
\bibitem [{\citenamefont {Freidel}\ \emph
  {et~al.}(2017{\natexlab{b}})\citenamefont {Freidel}, \citenamefont {Leigh},\
  and\ \citenamefont {Minic}}]{Freidel:2017wst}%
  \BibitemOpen
  \bibfield  {author} {\bibinfo {author} {\bibfnamefont {L.}~\bibnamefont
  {Freidel}}, \bibinfo {author} {\bibfnamefont {R.~G.}\ \bibnamefont {Leigh}},\
  and\ \bibinfo {author} {\bibfnamefont {D.}~\bibnamefont {Minic}},\ }\href
  {https://doi.org/10.1007/JHEP09(2017)060} {\bibfield  {journal} {\bibinfo
  {journal} {JHEP}\ }\textbf {\bibinfo {volume} {09}},\ \bibinfo {pages}
  {060}},\ \Eprint {https://arxiv.org/abs/1706.03305} {arXiv:1706.03305
  [hep-th]} \BibitemShut {NoStop}%
\bibitem [{\citenamefont {Freidel}\ \emph
  {et~al.}(2017{\natexlab{c}})\citenamefont {Freidel}, \citenamefont {Leigh},\
  and\ \citenamefont {Minic}}]{Freidel:2017nhg}%
  \BibitemOpen
  \bibfield  {author} {\bibinfo {author} {\bibfnamefont {L.}~\bibnamefont
  {Freidel}}, \bibinfo {author} {\bibfnamefont {R.~G.}\ \bibnamefont {Leigh}},\
  and\ \bibinfo {author} {\bibfnamefont {D.}~\bibnamefont {Minic}},\ }\href
  {https://doi.org/10.1103/PhysRevD.96.066003} {\bibfield  {journal} {\bibinfo
  {journal} {Phys. Rev. D}\ }\textbf {\bibinfo {volume} {96}},\ \bibinfo
  {pages} {066003} (\bibinfo {year} {2017}{\natexlab{c}})},\ \Eprint
  {https://arxiv.org/abs/1707.00312} {arXiv:1707.00312 [hep-th]} \BibitemShut
  {NoStop}%
\bibitem [{\citenamefont {Freidel}\ \emph
  {et~al.}(2019{\natexlab{b}})\citenamefont {Freidel}, \citenamefont
  {Kowalski-Glikman}, \citenamefont {Leigh},\ and\ \citenamefont
  {Minic}}]{Freidel:2018apz}%
  \BibitemOpen
  \bibfield  {author} {\bibinfo {author} {\bibfnamefont {L.}~\bibnamefont
  {Freidel}}, \bibinfo {author} {\bibfnamefont {J.}~\bibnamefont
  {Kowalski-Glikman}}, \bibinfo {author} {\bibfnamefont {R.~G.}\ \bibnamefont
  {Leigh}},\ and\ \bibinfo {author} {\bibfnamefont {D.}~\bibnamefont {Minic}},\
  }\href {https://doi.org/10.1103/PhysRevD.99.066011} {\bibfield  {journal}
  {\bibinfo  {journal} {Phys. Rev. D}\ }\textbf {\bibinfo {volume} {99}},\
  \bibinfo {pages} {066011} (\bibinfo {year} {2019}{\natexlab{b}})},\ \Eprint
  {https://arxiv.org/abs/1812.10821} {arXiv:1812.10821 [hep-th]} \BibitemShut
  {NoStop}%
\bibitem [{\citenamefont {Raffelt}(2011)}]{Raffelt:2011yb}%
  \BibitemOpen
  \bibfield  {author} {\bibinfo {author} {\bibfnamefont {G.~G.}\ \bibnamefont
  {Raffelt}},\ }\href {https://doi.org/10.1103/PhysRevD.83.105022} {\bibfield
  {journal} {\bibinfo  {journal} {Phys. Rev. D}\ }\textbf {\bibinfo {volume}
  {83}},\ \bibinfo {pages} {105022} (\bibinfo {year} {2011})},\ \Eprint
  {https://arxiv.org/abs/1103.2891} {arXiv:1103.2891 [hep-ph]} \BibitemShut
  {NoStop}%
\bibitem [{\citenamefont {Pehlivan}\ \emph {et~al.}(2011)\citenamefont
  {Pehlivan}, \citenamefont {Balantekin}, \citenamefont {Kajino},\ and\
  \citenamefont {Yoshida}}]{Pehlivan:2011hp}%
  \BibitemOpen
  \bibfield  {author} {\bibinfo {author} {\bibfnamefont {Y.}~\bibnamefont
  {Pehlivan}}, \bibinfo {author} {\bibfnamefont {A.}~\bibnamefont
  {Balantekin}}, \bibinfo {author} {\bibfnamefont {T.}~\bibnamefont {Kajino}},\
  and\ \bibinfo {author} {\bibfnamefont {T.}~\bibnamefont {Yoshida}},\ }\href
  {https://doi.org/10.1103/PhysRevD.84.065008} {\bibfield  {journal} {\bibinfo
  {journal} {Phys. Rev. D}\ }\textbf {\bibinfo {volume} {84}},\ \bibinfo
  {pages} {065008} (\bibinfo {year} {2011})},\ \Eprint
  {https://arxiv.org/abs/1105.1182} {arXiv:1105.1182 [astro-ph.CO]}
  \BibitemShut {NoStop}%
\end{thebibliography}%

\end{document}